\documentclass[article,twocolumn,showpacs,superscriptaddress,amsmath,amssymb]{revtex4}
\usepackage{cancel}
\usepackage{maybemath}
\usepackage{xcolor}
\usepackage{graphicx}

\begin{document}
Published as:\\R. Ehrlich. Advances in Astronomy, Vol. 2019, 2820492\\
\title{Review of the empirical evidence for superluminal particles\\and the $3+3$ model of the neutrino masses}
% repeat the \author .. \affiliation  etc. as needed
% \email, \thanks, \homepage, \altaffiliation all apply to the current
% author. Explanatory text should go in the []'s, actual e-mail
% address or url should go in the {}'s for \email and \homepage.
% Please use the appropriate macro foreach each type of information
% \affiliation command applies to all authors since the last
% \affiliation command. The \affiliation command should follow the
% other information
% \affiliation can be followed by \email, \homepage, \thanks as well.
\author{Robert Ehrlich}
\affiliation{George Mason University, Fairfax, VA 22030}
\email{rehrlich@gmu.edu}
%Collaboration name if desired (requires use of superscriptaddress
%option in \documentclass). \noaffiliation is required (may also be
%used with the \author command).
%\collaboration can be followed by \email, \homepage, \thanks as well.
%\collaboration{}
%\noaffiliation
\date{\today}
\begin{abstract}
A review is given of hypothetical faster-than-light tachyons and the development of the author's $3+3$ model of the neutrino mass states, which includes one tachyonic mass state doublet.  Published empirical evidence for the model is summarized, including an interpretation of the mysterious Mont Blanc neutrino burst from SN 1987A as being due to tachyonic neutrinos having $m^2=-0.38 eV^2.$  This possibility  requires an 8 MeV antineutrino line from SN 1987A, which a new dark matter model has been found to support.  Furthermore, this dark matter model is supported by several data sets: $\gamma-$rays from the galactic center, and the Kamiokande-II neutrino data on the day of SN 1987A.  The KATRIN experiment should serve as the unambiguous test of the $3+3$ model and its tachyonic mass state.
\end{abstract}
% insert suggested PACS numbers in braces on next line
 % insert suggested keywords - APS authors don't need to do this
%\keywords{neutrino, neutrino mass, tritium beta decay, KATRIN experiment}
%\maketitle must follow title, authors, abstract, \pacs, and \keywords
\maketitle
% body of paper here - Use proper section commands
% References should be done using the \cite, \ref, and \label commands
% switch to two-column layout

\section{v $>$ c Tachyons}
Hypothetical faster-than-light particles, now known as tachyons, were first suggested in 1962 by Bilaniuk, Deshpande, and Sudarshan as a way to extend special relativity to the $v>c$ realm.~\citep{Bi1962}  Sudarshan and colleagues noted that if a particle was allowed to have a rest mass that was imaginary, or $m^2<0$ one could use the usual formula to compute its real total energy $E=mc^2/\sqrt{1-v^2/c^2},$ as long as the particle was never allowed to have $v<c.$  For those concerned about the meaning of an imaginary rest mass, Ref.~\citep{Bi1962} reminds us that only energy and momentum, by virtue of their direct observability and conservation in interactions, must be real and that the hypothetical imaginary rest mass particles offend only the traditional way of thinking.  In this scheme $v=c$ becomes a two-way infinite energy barrier -- an upper limit to normal ($m^2>0$) particles and a lower limit to hypothetical tachyons, thus allowing all matter to be divided into three classes with $m^2$ being positive, negative or zero.  Moreover, tachyons have the weird property as Fig. 1 shows of speeding up as they lose energy, and approaching infinite speed as E approaches zero.  There are, of course, cases of allowed superluminal motion.  Thus, Recami and others have considered localized X-shaped solutions to Maxwell's equations,~\citep{Re1998}, quantum tunneling through two successive barriers,~\citep{Re2002} and the apparent separation speed of quasars~\citep{Ba1989}.   A nice overview of these and other allowed types of superluminal motion can be found in Recami~\citep{Re2001, Re2008}.  However, in these cases there is no superluminal motion of particles or information with the possibility of a violation of causality, making them outside the scope of this review.  

Since the original tachyon paper~\citep{Bi1962} Recami and Mignani,~\citep{Re1974}, Recami~\citep{Re1986} and later Cohen and Glashow~\citep{Co2011} and other theorists have suggested various ways to accommodate $v>c$ particles, including the adoption of nonstandard dispersion relations, which can avoid imaginary rest masses, but at the price (in the Cohen-Glashow case) of making the value of a particle's rest mass dependent the choice of reference frame. 
\begin{figure}
\centerline{\includegraphics[angle=0,width=1.0\linewidth]{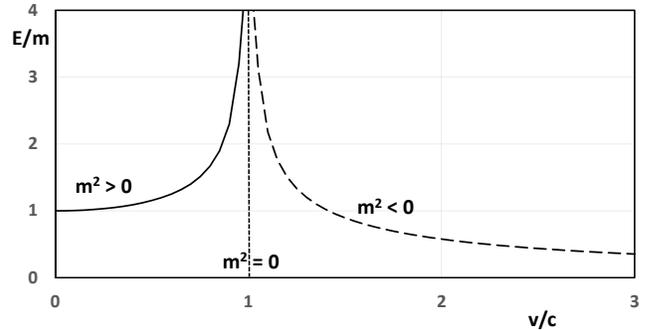}}
\caption{$E/|m|$ versus v/c for $m^2>0, m^2<0$ and $m^2=0$ particles.}
\end{figure}
$v>c$ or ``classical" tachyons are not taken seriously by most physicists because of their obnoxious theoretical properties, and the repeated failed attempts to find unambiguous evidence for their existence.  These attempts include some well-known mistaken claims, most recently by the OPERA Collaboration in 2011.~\citep{Ad2011}  In fairness to OPERA, the initial paper made no discovery claim, and it merely announced their observed $v>c$ anomaly with the intent of promoting further inquiry and debate.  As is well-known the group later found several experimental flaws and their corrected neutrino speed was consistent with c.~\citep{Ad2012}  In any case, most theoretical (and experimental) physicists have little use for the faster-than-light variety of tachyon, which has been considered a violation of relativity (Lorentz Invariance) and Causality (prohibition against backward-time signalling), although it is also true that some theorists have postulated ways around such difficulties by for example postulating a preferred reference frame or small violations of Lorentz Invariance.~\citep{Re1987,Ci1999,Ra2010}  Moreover, while most physicists abhor the $v>c$ classical  tachyon they have much greater affinity for another variety that is widely used in field theory.~\citep{Se2002}  These more reputable tachyons have imaginary mass quanta, but no $v>c$ propagation speed, the field associated with the Higgs particle being the best known example.  In particular, the  imaginary mass quanta of the Higgs field cause instabilities leading to a spontaneous decay or condensation, but again no $v>c$ propagation.  In the rest of this paper the word tachyon refers to the $v>c$ and $m^2<0$ ``disreputable" variety.

Given the current state of experimental physics, the only known particle that could be a tachyon is one of the neutrinos, a possibility raised by Chodos, Kostelecky and Hauser in a 1985 paper.~\citep{Ch1985}  Since the neutrino's observed mass is so close to zero, we cannot be certain yet whether $m_\nu^2>0$ or $m_\nu^2<0,$ although it is known that $m_\nu^2\ne 0,$ for at least some neutrinos in order for neutrino oscillations to be possible -- a connection that was explored in a 1986 paper by Giannetto et al.~\citep{Gi1986}.  Considering the two types of measurements, $v$ or $m^2,$ it is the latter that permits us to put much tighter constraints on whether the neutrino is or is not a tachyon.  Thus, if neutrinos in fact had a velocity that was slightly in excess of c by an amount say half the present experimental uncertainty, then their computed $|m^2|$ would need to be orders of magnitude above what would have been readily observed by now in direct mass experiments.  

\section{Direct neutrino mass experiments}
The most common ``direct" (model independent) method of measuring the neutrino (or antineutrino) mass is to look for distortions of the $\beta-$decay spectrum near its endpoint.   In these experiments an antineutrino is emitted is in the electron flavor state $\nu_e$ which is a quantum mechanical mixture of states $\nu_j$ having specific masses $m_j$ with weights $U_{ej}, $ i.e., $\nu_e=\sum U_{ej}\nu_j.$  In general, if one can ignore final state distributions, the phase space term describes the spectrum fairly well near the endpoint $E_0,$ and it can be expressed in terms of the effective electron neutrino mass using the square of the Kurie function.  

\begin{equation}
K^2(E)=(E_0-E)\sqrt{(E_0-E)^2-m^2_\nu\rm{(eff)}}
\end{equation}

In Eq.1 the $\nu_e$ effective mass is defined in single $\beta-$decay by this weighted average of the individual $m_j^2$:
\begin{equation}
m^2_\nu\rm{(eff)}=\sum |U_{ej}|^2 m_j^2
\end{equation}

However, if the individual $m_j$ could be distinguished experimentally, one would need to use a weighted sum of spectra for each of the $m_j$ with weights $|U_{ej}|^2$~\citep{Gi2007}

\begin{equation}
K^2(E)=(E_0-E)\sum |U_{ej}|^2\sqrt{(E_0-E)^2-m_j^2}
\end{equation}

Note that when $(E_0-E)^2-m_j^2$ is negative it is replaced by zero in Eq. 1 and 3 so as to avoid negative values under the square root.  Given the form of Eq. 1 a massless neutrino yields a quadratic result: $K^2(E)=(E_0-E)^2$ near the endpoint, while a neutrino having an effective $\nu_e$ mass $m^2_\nu\rm({eff})>0$ would result in the spectrum ending a distance $m_\nu\rm({eff})$ from the endpoint defined by the decay Q-value.  Moreover using Eq. 3 in the case of $m_j^2>0$ neutrinos of distinguishable mass, we would find that the spectrum shows kinks for each mass at a distances $m_j$ from the endpoint defined by the decay Q-value.   These direct mass experiments are extraordinarily difficult in light of systematic effects that also distort the spectrun, and the very small number of electrons observed near the spectrum endpoint.  As of October 2018 they have only set upper limits on $m_\nu\rm{(eff)}<2eV,$~\citep{Pa2016} at least according to conventional wisdom.  The possibility of observing a $m^2_\nu<0$ neutrino in direct mass experiments is discussed later, but for now we merely note that the results of nearly all such experiments that have in fact found best fit $m_\nu^2<0$ values should not be taken at face value.  If they are not due to systematic errors, these results have a simple explanation within the $3+3$ model, as will be discussed later.

Some experiments hope to look for massive (sterile) neutrinos in oscillation experiments, but these experiments do not measure neutrino masses directly, but rather differences in the $m^2$ values of the states making up an oscillating pair, so they would not be sensitive to whether one or both of those states have $m^2>0$ or $m^2<0.$   One could however possibly observe $m_\nu^2<0$ neutrinos from a galactic supernova.  Any neutrinos having $v>c$ would of course arrive earlier than those having $m^2>0,$ assuming they all started out approximately simultaneously, and those having higher energy would arrive {\emph{later}} than those with lower energy.

\section{SN 1987A and the Mont Blanc burst}
Galactic supernovae are quite rare, occurring an estimated $2\pm 1$ times per century, making SN 1987A a precious treasure that has deserved the very careful attention it has received, with thousands of papers written about it to date.  Although many supernovae have now been observed in other galaxies, only SN 1987A was close enough to study the neutrinos it produced during the final collapse of the core of its progenitor star.  In fact four neutrino detectors then operating (see Fig. 2 caption) each observed a burst lasting 5-15 seconds, representing a mere 30 neutrinos (or antineutrinos) in total.  Three of the bursts occurred within a matter of seconds of each other as expected, but the fourth detector located under Mont Blanc detected its burst of 5 neutrino events almost 5 hours (16,900 sec) earlier than the others.~\citep{Ag1987,Ag1988}  As a result, most physicists with a few notable exceptions,~\citep{Br1992,Fr2015,Gi1999} have chosen to dismiss the Mont Blanc burst as having nothing to do with SN 1978A.

\subsection{Using SN 1987A to find the neutrino mass}

SN 1987A like all supernovae create huge numbers of neutrinos and antineutrinos having all three flavors, e, $\mu,$ and $\tau,$ but it is the electron flavor that is typically detected.  Conventional wisdom has it that SN 1987A was only able to set an upper limit on the mass of the electron neutrino $\nu_e$ of 5.7 eV~\citep{Pa2016} or more depending on how the analysis is done, and that there was no hint of a $m^2<0$ mass state, assuming one ignores the Mont Blanc burst.  As in the case of direct mass experiments, these standard analyses assume that the separate active mass states comprising the electron neutrino are so close in mass that one can only hope to observe a single effective mass for $\nu_e.$   This assumption of only a single effective mass being observable is supported by the standard model of three active neutrinos whose $m^2$ values are separated from one another by two very small quantities, the solar and atmospheric mass differences: $\Delta m^2_{sol} =7.53\times 10^{-5} eV^2$ and $\Delta m^2_{atm} =2.44\times 10^{-3} eV^2.$

\subsection{Origin of the $3+3$ model}
In a 2012 paper~\citep{Eh2012} the author analyzed the SN 1987A data without making the assumption that only a single effective mass could be found, and he rediscovered a most peculiar fact that Huzita\citep{Hu1987} and Cowsik~\citep{Co1988} had pointed out soon after SN 1987A was observed.   If one assumes near-simultaneous emissions then all the 25 neutrinos (ignoring the five from Mont Blanc) were consistent with having one of two outlandishly large masses, $m_1=4.0\pm 0.5$ eV and $m_2=21.4\pm 1.2$ eV.~\citep{Eh2012}  This result depends on a kinematic relation between the $i^{th}$ neutrino energy $E_i$ and its travel time $t_i$ (relative to a photon) which in the limit $E_i>>m_i$ can be written:  

\begin{equation}
\frac{1}{E_i^2} = \left(\frac{2}{Tm_i^2}\right)t_i
\end{equation}

Here T is the travel time of a photon (around 168 ky), and $m_i$ is the mass of the $i^{th}$ observed neutrino, and $t_i>0$ $(t_i<0)$ means slower (faster) than light.  Essentially, Eq. 4 requires that neutrinos having a given mass $m_i$ should lie on or near a straight line in a plot of $1/E^2$ versus time $t_i$ whose slope reveals the value of $m_i^2.$  Thus, the question of whether the neutrinos are consistent with a single mass is left up to the data to answer, which as can be seen from Fig. 2 taken from Ref.~\citep{Eh2012} would seem to favor two separate $m^2>0$ masses and not one effective mass.   Regarding the assumption of simultaneous emissions Ref.~\citep{Eh2012} argues that most or all of the observed neutrinos from SN 1987A were emitted within an interval of $\pm0.2s.$  Moreover, given the usual choice of $t=0$ for the first arriving neutrino in each of the unsynchronized detectors (except Mont Blanc) there are probably $\pm1$ s implied horizontal error bars for the points in Fig. 2.

\begin{figure}
\centerline{\includegraphics[angle=0,width=0.8\linewidth]{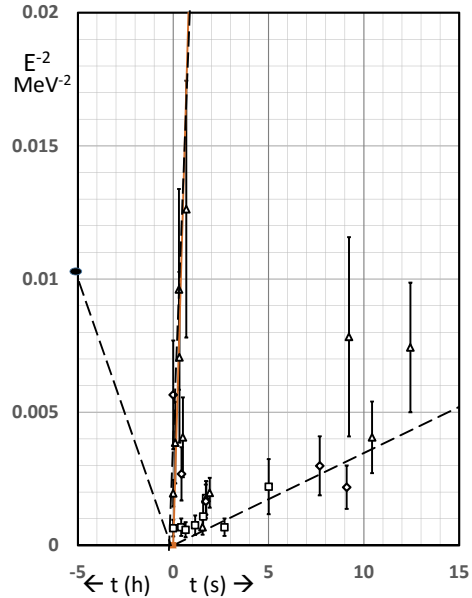}}
\caption{\label{fig3}A plot of $1/E_\nu^2$ versus observation time for the events seen in the four neutrino detectors operating at the time of SN 1987A:  open diamonds (5 Baksan events), open squares (12 Kamiokande-II events), open triangles (8 IMB events).  The single dot near t = - 5h shown without error bars represents the 5 Mont Blanc events.  The dashed positive sloped lines correspond to masses $m_1=4.0$ eV, and $m_2=21.4$ eV according to Eq. 2.  The $3+3$ model called for a third mass $m_3$ that was a tachyon, but it initially explicitly rejected the idea that the 5 Mont Blanc events defined one, as discussed in the text.}
\end{figure}

The $3+3$ model was proposed in 2013~\citep{Eh2013} based on the anomalously large values for $m_1$ and $m_2$ that are implied by Fig. 2.  Clearly, the only way to accommodate the very small $\Delta m^2_{sol}$ and $\Delta m^2_{atm}$ was to assume that $m_1$ and $m_2$ were each active-sterile doublets -- see Fig. 3.  This model is in marked contrast to the $(3 + 0)$ conventional model lacking sterile neutrinos which is described by three mixing angles.  If any sterile neutrinos are assumed to exist within the conventional framework they must mix very little with the three active neutrino states so as to preserve unitarity of the $U_{ij}$ matrix.  Furthermore, for any model (like $3+3$) that has two active-sterile doublets each with large mixing there must of course be a third doublet, since it is well established that there are three active states.  In such a model there are a total of 15 mixing angles and 10 phases, making it much more complex to describe oscillation phenomena than the conventional model.  

\subsection{The $m^2<0$ doublet in the $3+3$ model}
In an earlier paper~\citep{Eh2015} the author had suggested a tachyonic value for the $\nu_e$ effective flavor state mass, i.e., $m^2_\nu\rm{(eff)}=-0.11\pm 0.02eV^2$ and for this to be the case Eq. 2  would require $m_3^2<0.$  The particular choice $m^2_3 \approx -0.2 keV^2$ was made after a remarkable numerical coincidence was discovered, namely that with the pair of doublet splittings $\Delta m_1^2=\Delta m^2_{sol}$ and $\Delta m_2^2=\Delta m^2_{atm},$ one finds identical fractional splittings $\Delta m_1^2/m_1^2=\Delta m_2^2/m_2^2$ for the two doublets.  The choice of the 3rd doublet mass then became obvious.  Given that short baseline experiments have suggested an oscillation having $\Delta m^2_{sbl}\approx 1 eV^2$ if one chose $m_3^2\approx -0.2keV^2,$ all three doublets would then have identical fractional splittings, i.e., 
\begin{equation}
\frac{\Delta m^2_1}{m_1^2}=\frac{\Delta m^2_2}{m_2^2}=
\frac{\Delta m^2_3}{m_3^2}
\end{equation}

\begin{figure}
\centerline{\includegraphics[angle=0,width=0.8\linewidth]{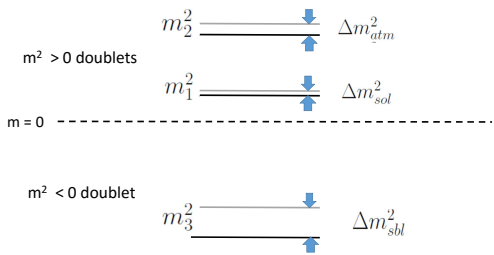}}
\caption{The three active-sterile doublets and their splittings in the $3+3$ model (not drawn to scale).  Note that two doublets have $m^2>0$ and one has $m^2<0.$  The values for the three masses found from a non-standard analysis of SN 1987A data are given in the text.}
\end{figure}

\subsection{Support for the $3+3$ model}
A model as speculative as $3+3$ especially considering its $m_3^2<0$ doublet clearly needs empirical support before it deserves to be taken seriously.  Previous papers~\citep{Ch2014,Eh2016} have in fact provided such support, which is very briefly summarized here.  First it was shown that the dark matter radial distribution in the Milky Way Galaxy could be fit using a nearly degenerate gas of neutrinos having a mass very close to 21.4 eV, and that for clusters of galaxies the dark matter distributions could be fit using neutrinos having a 4.0 eV mass~\citep{Ch2014} -- these being the $m^2>0$ masses in the $3+3$ model.  The $m^2<0$ mass would not be associated with dark matter, but rather with dark energy, as suggested by various authors.~\citep{Ba2003, Sc2018}.  More recently, it has been shown that fits to the $\beta-$spectrum near its endpoint for the three most precise pre-KATRIN tritium $\beta$-decay experiments (by the Mainz, Troitsk and Livermore Collaborations) could be achieved using the three masses in the $3+3$ model, and moreover these fits were significantly better than the fit to a single effective mass, which only gives an upper limit $m_\nu\rm{(eff)}<2 eV$ for the $\bar{\nu}_e$ mass.~\citep{Eh2016}  

The most prominent spectral feature in the $3+3$ model is a kink 21.4 eV before the endpoint, which appears in the data from all three experiments -- see Figs. 4, 5 and 6 taken from Ref.~\citep{Eh2016}.   
The evidence for this kink in the Mainz data rests on a single data point in their 1998-99 data that is $5\sigma$ above the $m=0$ curve, for reasons explained in the caption to Fig. 4.  The Troitsk spectrum published 1999 (Fig. 5) clearly shows the kink at the up arrow in Fig. 5.~\citep{Lo1999}  However, the location of that kink agrees well with the $3+3$ model fit (the solid curve added by the author) only after an adjustment is made to the scale of the energy axis.  Such an adjustment to the data (moving the kink from 10 to 20 eV before the endpoint) might seem unwarranted were it not actually called for in the most recent 2012 Troitsk publication -- see Fig. 7 in Ref.~\citep{As2012}.   In this newer analysis, the Troitsk authors `` did not employ overly short runs and runs in which external parameters have large uncertainties."  As a result of this elimination of some runs, they have withdrawn any claim of statistical significance of their unexplained anomaly.  Unfortunately in that 2012 reassessment the authors have chosen no longer to display the spectrum, which is why the originally published Troitsk spectrum was used in our Fig. 5.  Moreover, despite their withdrawal of a claim of statistical significance, some evidence for a kink clearly must remain in their data, because as shown by the dashed and solid horizontal lines in Fig. 8 in Ref.~\citep{As2012} the amplitude of the kink averaged over all runs is about 3/4 that in the original spectrum.

\begin{figure}
\centerline{\includegraphics[angle=0,width=1.0\linewidth]{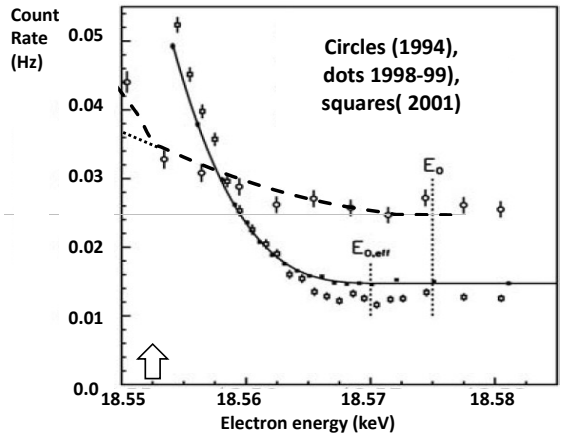}}
\caption{The published data for the Mainz Collaboration taken during different years as it appeared in ref.~\citep{Eh2016}. The solid curve was their $m = 0$ fit to the 1998-99 data. The dashed $3+3$ curve has been added after adjusting the background level and the vertical scale so as to fit the 1994 data. The Mainz data from later years do not show the predicted kink (at the location of the up arrow), because they Mainz did not publish their data beyond 20 eV from the spectrum endpoint in those years.}
\end{figure}

\begin{figure}
\centerline{\includegraphics[angle=0,width=1.0\linewidth]{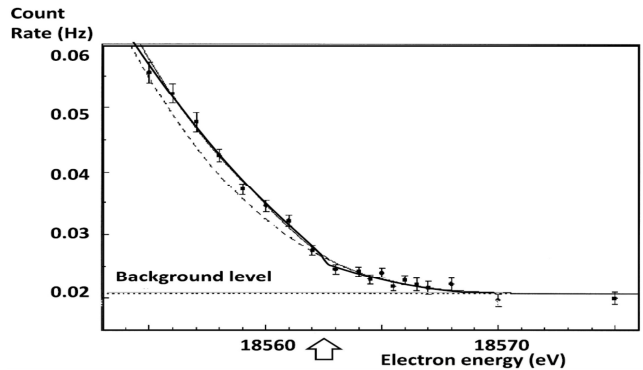}}
\caption{The published spectrum for the Troitsk Collaboration taken from Ref.~\citep{Lo1999}. The solid curve is this authors fit to the $3+3$ model from Ref.\citep{Eh2016} after an adjustment to the energy axis, which fits the data much better than the standard quadratic (dashed curve) for the zero mass case.  The most recent (2012) Troitsk reassessment of their anomaly no longer regards it as statistically significant as discussed in the text and in more detail in Ref.~\citep{Eh2016}.}
\end{figure}

The Livermore Collaboration also has chosen not to display the spectrum itself but instead the residuals from a fit to the data using the standard $m_\nu\rm{(eff)}=0$ expected spectrum.  When only residuals are plotted the kink predicted by the $3+3$ model shows up not as a kink but instead as a spurious spectral line broadened by resolution near the endpoint or alternatively a best fit value for $m^2_\nu\rm{(eff)}$ which is negative -- see Fig. 6.  Thus, as noted earlier, the $3+3$ model can account for the artifactual $m^2_\nu\rm{(eff)}<0$ fitted value found in nearly all direct mass experiments.\\
\begin{figure}

\centerline{\includegraphics[angle=0,width=1.0\linewidth]{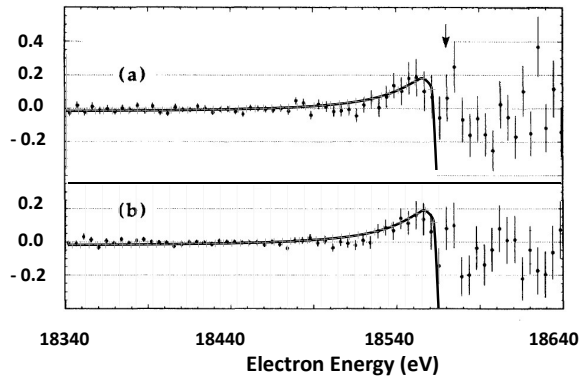}}
\caption{The published data for the Livermore Collaboration shown as residuals to a fit to the spectrum assuming zero effective mass, taken from Ref.~\citep{Eh2016}.  The curve added by the author shows what the residual graph would look like assuming the true spectrum was described by the $3+3$ model masses.  (a) and (b) are for two subsamples of the data, and the arrow shows the spectrum endpoint.}
\end{figure}

\subsection{Evidence for the $m^2<0$ mass}
It has been shown in the previous section (and in more detail in Ref.~\citep{Eh2016}) that the $3+3$ model gives better fits than the conventional $m_\nu \rm{(eff)}=0$ parabolic curve to three tritium beta decay experiments.  However, given the sparcity of data near the endpoint it has only been the most prominent feature of the model (the kink in the spectrum 21.4 eV before the endpoint) that the data were able to reveal.  A recent paper has provided evidence for the tachyonic mass in the model by showing how it explains the mysterious Mont Blanc neutrino burst seen on the date of SN 1987A,~\citep{Eh2017} a possibility first raised by Giani.~\citep{Gi1999} This burst has been a mystery, not only because of its early arrival, but also because the 5 neutrinos comprising it have virtually the same energy (8 MeV) within measurement uncertainties -- something no model has previously explained.~\citep{Vi2015}  

Using Eq. 4, one can deduce a value for the tachyonic mass for the burst.  Thus, with $\Delta t = 16,900$ s and $E_{avg}=8.0$ MeV, Eq. 4 yields $m_{avg}^2 = -0.38$ keV$^2$ -- a mass value that is within a factor of two of the originally hypothesized value $m^2 \approx -0.2$ keV$^2.$~\citep{Eh2013}   One should not expect any better agreement because the mass value in the $3+3$ model was based on the estimated $\Delta m^2_{sbl}\approx 1$ eV$^2$ for the large $\Delta m^2$ oscillation claimed in short baseline experiments, which in fact is uncertain by over a factor of two.~\citep{Gi2013}  

It should be noted that the author initially ignored the Mont Blanc burst, assuming it could not be evidence for a tachyon.  The basis for that initial rejection follows from Eq. 4 and Fig. 2 which show that neutrinos having a specific $m^2<0$ mass should lie on a negatively sloped line and be distributed over the 5 h before $t=0.$  Thus, given the brief (7 sec) time interval of the 5 neutrinos making up the burst the only way they could correspond to a specific mass state would be for them to have almost exactly the same energy.   More specifically, the constancy of the energy of the 5 neutrinos needs to be (by Eq. 4) constant to a precision of $\Delta E/E_{avg}=7s/(2\times 16,900)=0.02 \%,$ which is essentially a line in the (anti)neutrino spectrum.   For the tachyonic interpretation of the Mont Blanc burst to be remotely plausible there needs to be some model of a core collapse supernova that gives rise to monochromatic 8 MeV neutrinos.

\section{Model for an 8 MeV neutrino line}
Neutrino lines are known to exist in the solar spectrum, but until now no existing model of core collapse supernovae has such a feature.  In a recent paper~\citep{Eh2017} the author has proposed a new supernova model for such a 8 MeV $\bar{\nu}$ line that invokes dark matter X particles of mass 8.4 MeV.  This particular mass  for dark matter is based on recently discovered isoscalar gauge bosons of mass $m_{Z'}=16.7\pm 0.6 MeV.$   These Z' particles have been postulated as carriers of a fifth force which could serve as a mediator between dark matter particles and standard model leptons.~\citep{Fe2016, Ch2016}  Thus, were one to postulate cold dark matter X particles of mass $m_X=m_{Z'}/2=8.4 MeV$ that annihilate they would yield monochromatic $8.4\pm 0.3$ MeV $\nu,\bar{\nu}$ pairs via the reaction $XX\rightarrow Z' \rightarrow \nu\bar{\nu},$ just as required.  Given that supernovae in our galaxy are quite rare, how could such a model be tested without waiting or the next one?
This $Z'$ mediated reaction model would apply not just to supernovae but also to any place with abundant dark matter, and sufficiently high temperature, such as the the galactic center.  Furthermore it is shown in Ref.~\citep{Eh2017} that the dark matter model predicts successfully three observed properties of the of MeV $\gamma-$rays from the galactic center -- see Table I and Fig. 7.   
\begin{figure}[t!]
\centerline{\includegraphics[angle=0,width=1.1\linewidth]{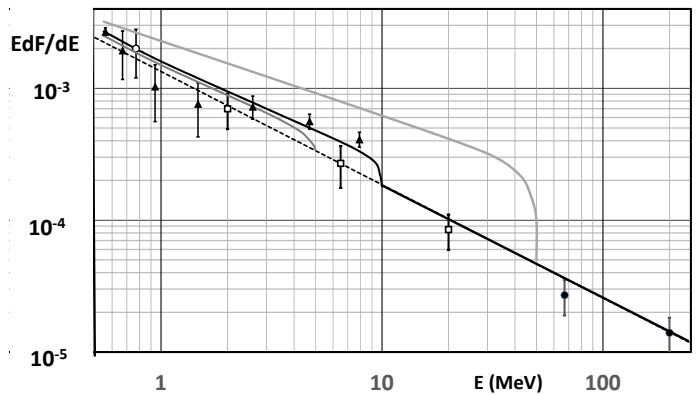}}
\vspace{-0.35in}
\caption{Spectrum, i.e., $E\times\frac{dF}{dE} (cm^{-2}s^{-1})$ versus energy for $\gamma-$rays from the inner galaxy for $E>511$ keV, as measured by 4 instruments: SPI(open circle), COMPTEL (open squares), EGRET (filled circles), and OSSE (filled triangles), as it appeared in ref.~\citep{Eh2017}.  All but the 7 OSSE points extracted from ref.~\citep{Ki2001} are from Prantzos et al.~\citep{Pr2010}, as are the 3 predicted enhancement curves above the straight line for positrons injected into a neutral medium at initial energies $E_0=5, 10, 50$ MeV displayed as the lower grey curve, the black curve, and the upper grey curve, respectively.  The sloped straight line (also from Ref.~\citep{Pr2010} is a power law fit to the spectrum at high and low energies.}
\end{figure}

\begin{figure}[t!]
\centerline{\includegraphics[angle=0,width=1.0\linewidth]{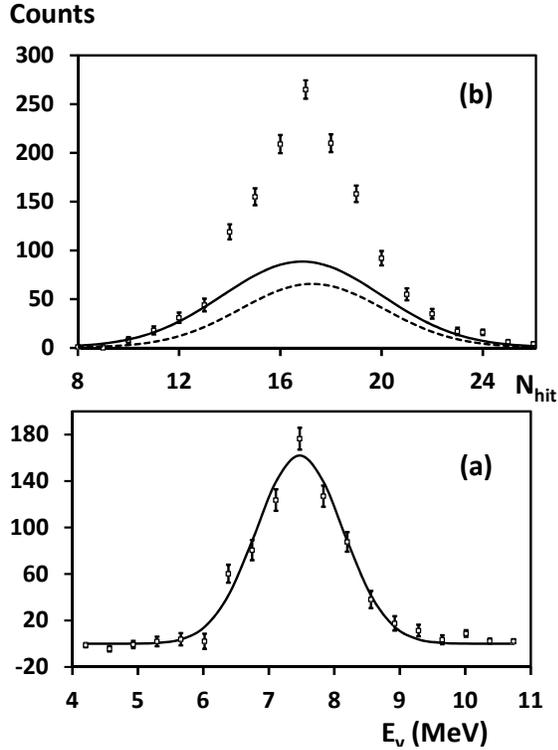}}
\caption{\label{fig3}Evidence for a neutrino line centered on 7.5 MeV in the Kamiokande-II data taken on the date of SN 1987A from Ref.~\citep{Eh2017}.  (b) shows a histogram of the raw data taken in the hours before and after the main burst, with $N_{hit}$ being a proxy for the neutrino energy, $E_\nu,$ and the solid and dashed curves being two versions of the background for the detector.  After subtracting the background (obtained from other K-II data) and converting the horizontal scale to neutrino energy one obtains (a) the background-subtracted curve.}
\end{figure}

Additional empirical support for an 8 MeV neutrino line (the basis of the dark matter model) is provided in ref.~\citep{Eh2017} based on Kamiokande-II data taken on the day of SN 1987A in the minutes and hours before and after the main 12-event burst in that detector.  These data are consistent with there being a antineutrino spectral line centered near 8 MeV that is broadened by $25\%$ energy resolution -- see Fig. 8.  The strength of the support for this claim of an 8 MeV line, of course, depends on the reliability of the background, which is discussed at length in ref.~\citep{Eh2017}.  

\begin{table}[h]
\centering
\begin{tabular}{lcc}
\hline\hline
Quantity &\hspace{0.25in} observed value &\hspace{0.25in} predicted value \\ 
\hline
$m_X$ & $10^{+5}_{-2}MeV$ & $ 8.4\pm 0.3 MeV$\\
$\sigma(\theta)$  & $2.5^0$   &  $2.4^0$\\
$T$                   & $10^3K$ & $10^3K$\\
\hline
\end{tabular}
\caption{Values of the dark matter particle mass $m_X,$ the temperature of the $\gamma-$ ray source T, and its angular radius $\sigma (\theta).$  The "observed" values are based on either direct observations or fits to the data, and the predictions follow from the $Z'_e/Z'_\nu$ mediated reaction model as discussed in Ref.\citep{Eh2017}.}
\end{table}

\begin{figure}[t!]
\centerline{\includegraphics[angle=0,width=0.8\linewidth]{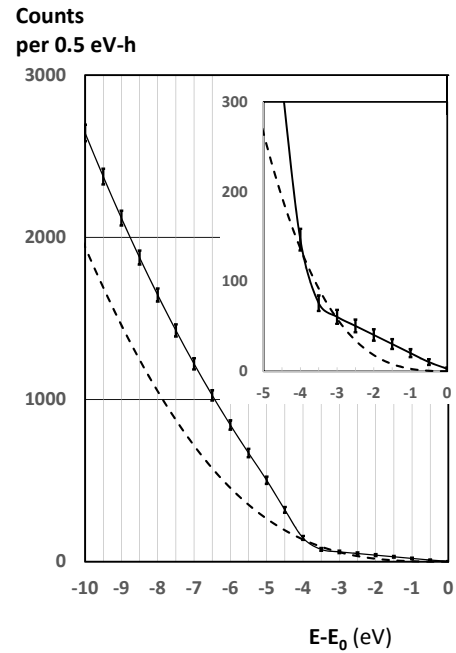}}
\caption{\label{fig3}Simulated KATRIN data based on the $3+3$ model masses for the last 10 eV of the spectrum as it appeared in ref.~\citep{Eh2017}.  The error bars are based on statistical uncertainties for one hour of data-taking for each of 24 energy bins of 0.5 eV width, with the event count normalized to yield the expected count rate at $E_0-E=20$ eV.  The dashed curve shows the all $m_j=0$ case after final state distributions and energy resolution have been included.  The insert shows the last 5 eV with an expanded vertical scale.}
\end{figure}

\section{The Mont Blanc burst again\label{Giani}}
Even though there were only 5 neutrinos in the anomalous Mont Blanc burst, and most physicists have dismissed this burst as being unrelated to SN 1987A, as the preceding sections have shown it played a very significant role in validating the $3 +3$ model with its $m^2<0$ mass state.  
Many speculative explanations have been suggested for the 5 hour early Mont Blanc neutrino burst,  assuming that it was not just a statistical fluctuation of the background.  These include (a) a double bang involving formation of a neutron star and followed by a black hole,~\citep{Br1992} (b) a new core collapse mechanism involving dark matter balls,~\citep{Fr2015} and (c) having the 5 neutrinos in the burst be a tachyon mass state, which was consistent with their equal energies $E\sim 8 MeV$ within the 15\% uncertainty~\citep{Gi1999}.  The tachyonic explanation of the Mont Blanc burst had until recently been regarded as extremely unlikely, even by this "tachyon enthusiast,"~\citep{Eh2013} given the lack of any known mechanism that would generate the required 8 MeV monochromatic neutrinos from a core collapse supernova.  A radical reassessment, however, is now warranted in favor of the tachyonic mass state explanation based on the author's 2016-2018 publications~\citep{Eh2016,Eh2017} which showed that: 
\begin{enumerate}
\item A new dark matter model explains why one would expect an 8 MeV monochromatic component of SN 1987A neutrinos
\item Empirical evidence supports that DM model based on the spectrum of galactic center $\gamma-$rays
\item Strong evidence ($S\sim30\sigma$) supports the existence of an 8 MeV line in the SN 1987A spectrum from the $N\sim 1000$ events in Kamiokande II on the day of SN 1987A
\item The value of the $m^2<0$ mass inferred from the Mont Blanc neutrinos is consistent with that originally postulated in the earlier $3+3$ model.
\item The $3+3$ model gives excellent fits to the three most accurate pre-KATRIN direct mass experiments.
\end{enumerate}
\section{The KATRIN experiment}

The KATRIN experiment~\citep{Dr2013} should prove or refute the existence of a tachyonic mass $m_3^2=-0.38 keV^2$ (and the two other masses in the $3+3$ model) in a short period of data-taking.  
Specifically, if the model is correct KATRIN should observe three features in the spectrum (associated with each of the three masses in the model).  The most prominent of these seen in Figs. 4-6 is the kink 21.4 eV before the endpoint.  Fig. 9 shows the two other predicted features:  a kink 4.0 eV before the endpoint due to the 4.0 eV mass state, and a linear decline in the last 4 eV, a feature based on the form of Eq. 3 when $m_3^2<0.$  The value used in generating this plot was the original mass in the model, $m_3^2\approx -0.2$ keV$^2$.  However, the interactive spreadsheet at Ref.~\citep{slider} allows the reader to see how these features in the spectrum change when one uses alternative masses including $m_3^2=-0.38$ keV$^2.$
Given that KATRIN has the sensitivity to see all three spectral features predicted by the $3+3$ model it should serve as an unambiguous test of the model's validity, including the tachyonic mass state.  However, even if the model should be proven incorrect KATRIN might still be consistent with a tachyonic flavor state, and in particular the earlier noted prediction by the author for the $\nu_e$ effective mass: $m^2_\nu\rm{(eff)}=-0.11\pm 0.02eV^2$\cite{Eh2015}.

The author declares that there is no conflict of interest regarding the publication of this paper.
\end{document}